\documentclass[twocolumn,trackchanges]{aastex6}

\usepackage{graphicx}
\usepackage{array}
\usepackage{amssymb}
\usepackage{natbib}
\usepackage{float}
\usepackage{color}
\usepackage[utf8]{inputenc}  
\usepackage{amsmath}  
\usepackage{comment}
\usepackage{footnote}
\usepackage{longtable}
\usepackage[caption2]{ccaption}
\usepackage[figuresright]{rotating}
\usepackage{booktabs}
\usepackage[flushleft]{threeparttable}

\newcommand{\degree}{\ensuremath{^\circ}}
\renewcommand*{\thefootnote}{\fnsymbol{footnote}}

\begin{document}


\title{Rare finding of a 100 kpc large double-lobed radio galaxy hosted in the Narrow line Seyfert 1 galaxy SDSS J103024.95+551622.7}



\author{Suvendu Rakshit\altaffilmark{1,2}, C. S. Stalin\altaffilmark{2}, Ananda Hota\altaffilmark{3} and Chiranjib Konar\altaffilmark{4}}

\altaffiltext{1}{Astronomy Program, Department of Physics and Astronomy, Seoul National University, Seoul 151-742; suvenduat@gmail.com}
\altaffiltext{2}{Indian Institute of Astrophysics, Block II, Koramangala, Bangalore-560034, India} 
\altaffiltext{3}{UM-DAE Centre for Excellence in Basic Sciences, Vidyanagari, Mumbai 400098, India}
\altaffiltext{4}{Amity Institute of Applied Sciences, Amity University Uttar Pradesh, Sector-125, NOIDA, U.P., India}

\begin{abstract}

Among the large varieties of active galactic nuclei (AGN) known, narrow line
Seyfert 1 (NLSy1) galaxies are a puzzling class, particularly after the
discovery of $\gamma$-ray emission in a handful of them using observations
from the {\it Fermi} Gamma-ray Space Telescope. Here, we report the discovery of a rare large double lobed radio source with its radio core associated with a
NLSy1 galaxy SDSS J103024.95+551622.7 at z = 0.435. 
The lobe separation is 116 kpc which is the second largest known projected size among NLSy1 radio sources. 
This finding is based on the analysis of
1.4 GHz data from
the Faint Images of the Radio Sky at Twenty-centimeters (FIRST) archives.
Along with the core and edge-brightened lobes we detected significant
(30\%) fraction of clear diffuse emission showing typical back-flow from
FR II radio galaxy lobes.
For the source, we estimated a jet power
of 3 $\times$ 10$^{44}$ erg s$^{-1}$ suggesting that its jet power is similar
to that of classical radio galaxies.  
Emission from the source 
is also found to be non-variable both in the optical and mid-infrared bands. 
Identification of more such sources may help to reveal new modes of AGN and 
understand their role in black hole galaxy evolution.

\end{abstract}




\section{Introduction}

Narrow Line Seyfert 1 (NLSy1) galaxies are a peculiar type of active 
galactic nuclei (AGN), characterized by the narrow H$\beta$ emission line with full width at half maximum (FWHM) less than 2000 km s$^{-1}$ and weak [O III] lines relative to H$\beta$ with
[O III]/H$\beta$ $<$ 3 \citep{1985ApJ...297..166O,1989ApJ...342..224G}. In addition, these sources 
have strong FeII emission in the UV optical region of the spectrum, soft X-ray 
excess, steep soft X-ray spectrum \citep{1996A&A...309...81W,1996A&A...305...53B} 
and show rapid large amplitude X-ray flux variations 
\citep{1999ApJS..125..317L,1996A&A...305...53B,2017MNRAS.466.3309R}.
These extreme properties shown 
by
NLSy1 galaxies are generally attributed to them having low mass
black holes(BHs) and accreting close to the Eddington limit 
\citep{1992ApJS...80..109B,2000ApJ...536L...5S}.
However, recent results on limited number of sources indicate that NLSy1 
galaxies too have BH masses similar to BLSy1 galaxies and their 
observed black hole mass deficit could be due to geometrical effects
\citep{2013MNRAS.431..210C,2016MNRAS.458L..69B,2016IJAA....6..166L,2017ApJ...842...96R}.

A fraction of about 7\% NLSy1 galaxies are known to be radio-loud characterized
by the radio loudness parameter $R$\footnote{R is defined as the ratio
of the flux density in the radio band at 5 GHz and the flux density in the optical $B$-band} $>$ 10  
\citep{1989AJ.....98.1195K}.  About 2\% of NLSy1 galaxies are found to be very 
radio-loud \citep[see][]
{2006AJ....132..531K,2006ApJS..166..128Z}. Increasing the sample of 
NLSy1 galaxies too yielded a low fraction of radio-loud NLSy1 (RL-NLSy1) galaxies 
of around 5\% \citep{2017ApJS..229...39R}. 
The low fraction of RL-NLSy1 galaxies
relative to the quasar population of AGN indicates that they are either
a rare population of sources or they are weak jet sources.  However, detection of strong radio emission, compact core with high brightness temperature, 
significant radio variability etc. in some NLSy1 galaxies indicate the presence of relativistic jets with blazar like characteristic \citep{2008ApJ...685..801Y,2017arXiv170310365L}. 
Some of the NLSy1 galaxies exhibit flat or inverted radio spectra while others 
show steep radio spectra \citep{2015ApJS..221....3G}. Interestingly, from about a dozen
RL-NLSy1 galaxies $\gamma$-ray emission  
has been detected by the  Large Area 
Telescope \citep[LAT; e.g.,][]{2009ApJ...699..976A,2015MNRAS.452..520D,2018ApJ...853L...2P} on board
the {\it Fermi} Gamma-ray Space Telescope,  making them 
very special candidates to study the jet formation process. The broad band
spectral energy distribution of these $\gamma$-ray detected NLSy1 galaxies 
have the double hump structure typical of blazars and specifically show 
resemblance to the flat spectrum category of AGN \citep{2018ApJ...853L...2P}. Accordingly, the low energy hump
of their SED is explained by synchrotron emission processes and their high energy hump is 
attributed to external Compton processes.

In terms of radio morphology, RL-NLSy1 galaxies are generally considered to be compact and unresolved
at the resolution of the images available in the Faint Images of the Radio Sky at Twenty-Centimeters
(FIRST) archives based on observations with the Very Large Array in its B configuration.
Only recently, RL-NLSy1 galaxies are known to have  kilo parsec (kpc) scale radio emission  
and as of today less than two  dozen NLSy1 galaxies have been detected with 
kpc scale jets  
\citep{2006AJ....131.1948W,2008A&A...490..583A,2010ApJ...717.1243G,2012ApJ...760...41D,
2015ApJ...798L..30D,2015ApJ...800L...8R,2017arXiv170403881C}, that too, with linear sizes lesser than 150 kpc.
As presented in Table \ref{table:obj}, we see that none of them show Fanaroff \& Riley II (FR II, \citealt{1974MNRAS.167P..31F})  
like double lobed structure at 100 kpc scale, 
to be comparable to standard radio galaxies \citep{1989AJ.....97....1K,1993MNRAS.263..139S}. Detection of 100 kpc scale emissions in RL-NLSy1 galaxies, making them 
comparable to standard radio galaxies is extremely important as this can shed 
new light into the jet-launching mechanism. Since radio jets are typically 
launched in low accretion rate on to high mass spinning black holes hosted in 
ellipticals, finding them in NLSy1 galaxies (typically low black hole mass, 
with high accretion rate in disk galaxies with gas rich central regions) is in 
clear contrary to expectations \citep{2018arXiv180103519B}.

In this paper, we report the detection of 100 kpc radio emission with standard FR II radio galaxy like lobes, from SDSS J103024.95+551622.7, 
which was recently classified as a NLSy1 galaxy by \citet{2017ApJS..229...39R} using Sloan Digital Sky Survey (SDSS) DR 12 spectroscopic data 
\citep{2015ApJS..219...12A}. This source has FWHM(H$\beta$)$=2170 \pm 27$ km s$^{-1}$, 
$F(\mathrm{O [III]})/F(\mathrm{H\beta}) =0.2$, $R_{4570}=0.15$, 
monochromatic luminosity at 5100 $\mathrm{\AA}$, $\log \lambda L_{5100} = 
45.13$ (erg s$^{-1}$) and a black hole mass of $\log M_{\rm {bh}}/M_{\odot} =7.96$ \citep{2017ApJS..229...39R}. In the soft X-ray (0.1–2 KeV) 
using ROSAT observations, it was found to have a steep photon index of $\Gamma=-2.61\pm 0.82$ \citep{2016A&A...588A.103B}. The paper is structured as follows. The sample and the data are discussed in section \ref{sec:data}. 
The results are given in section \ref{sec:results} followed by the interpretation 
and discussion in section \ref{sec:discussion}. The conclusions are given in section 
\ref{sec:conclusion}. A cosmology with $H_0 =70 \,\mathrm{km \,s^{-1}\, Mpc^{-1}}$,  $\Omega_m=0.3$ and $\Omega_{\lambda}=0.7$ is used throughout.

\begin{figure}
\centering
\resizebox{8.7cm}{7.5cm}{\includegraphics{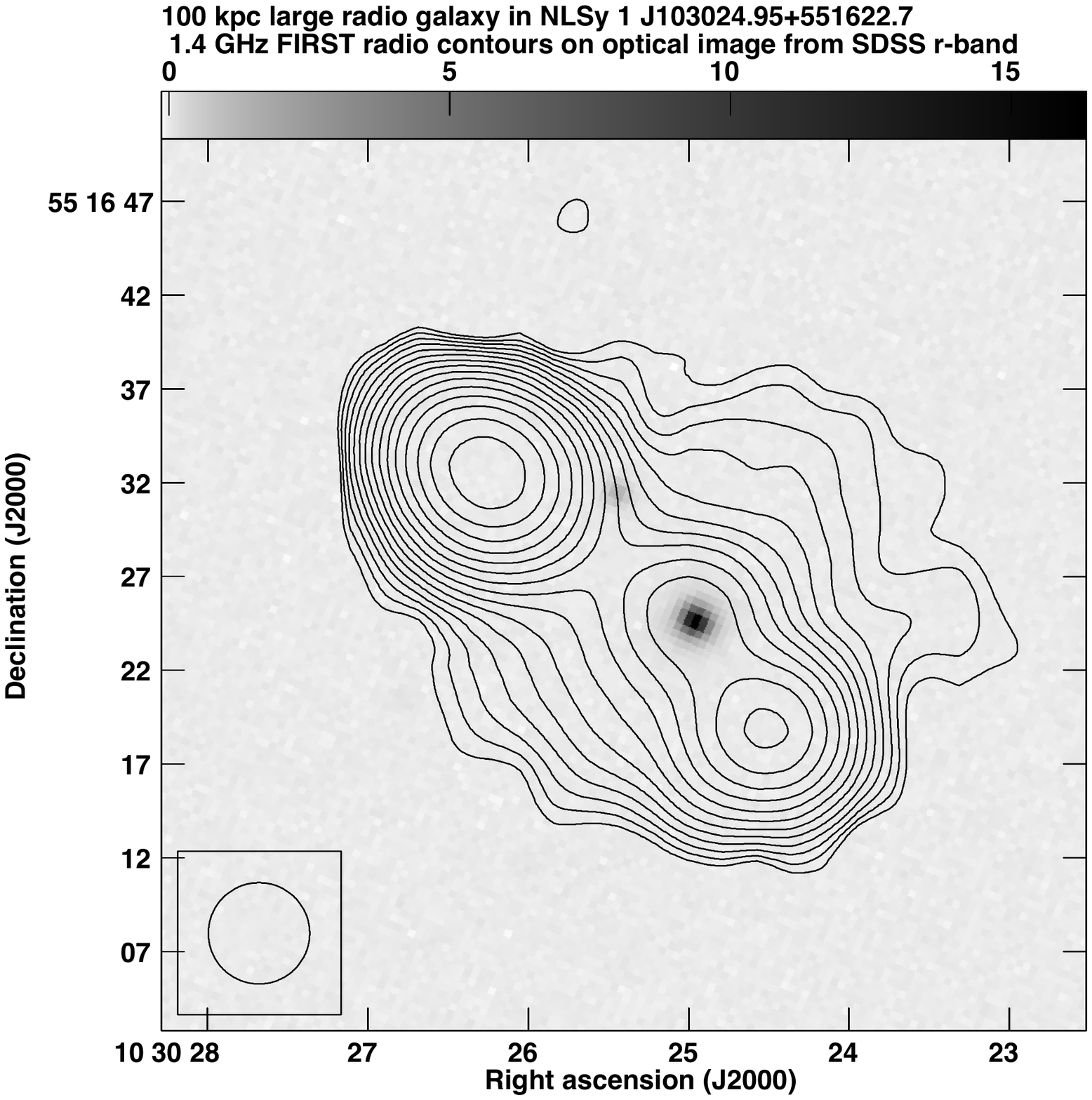}}
\resizebox{9cm}{5.6cm}{\includegraphics{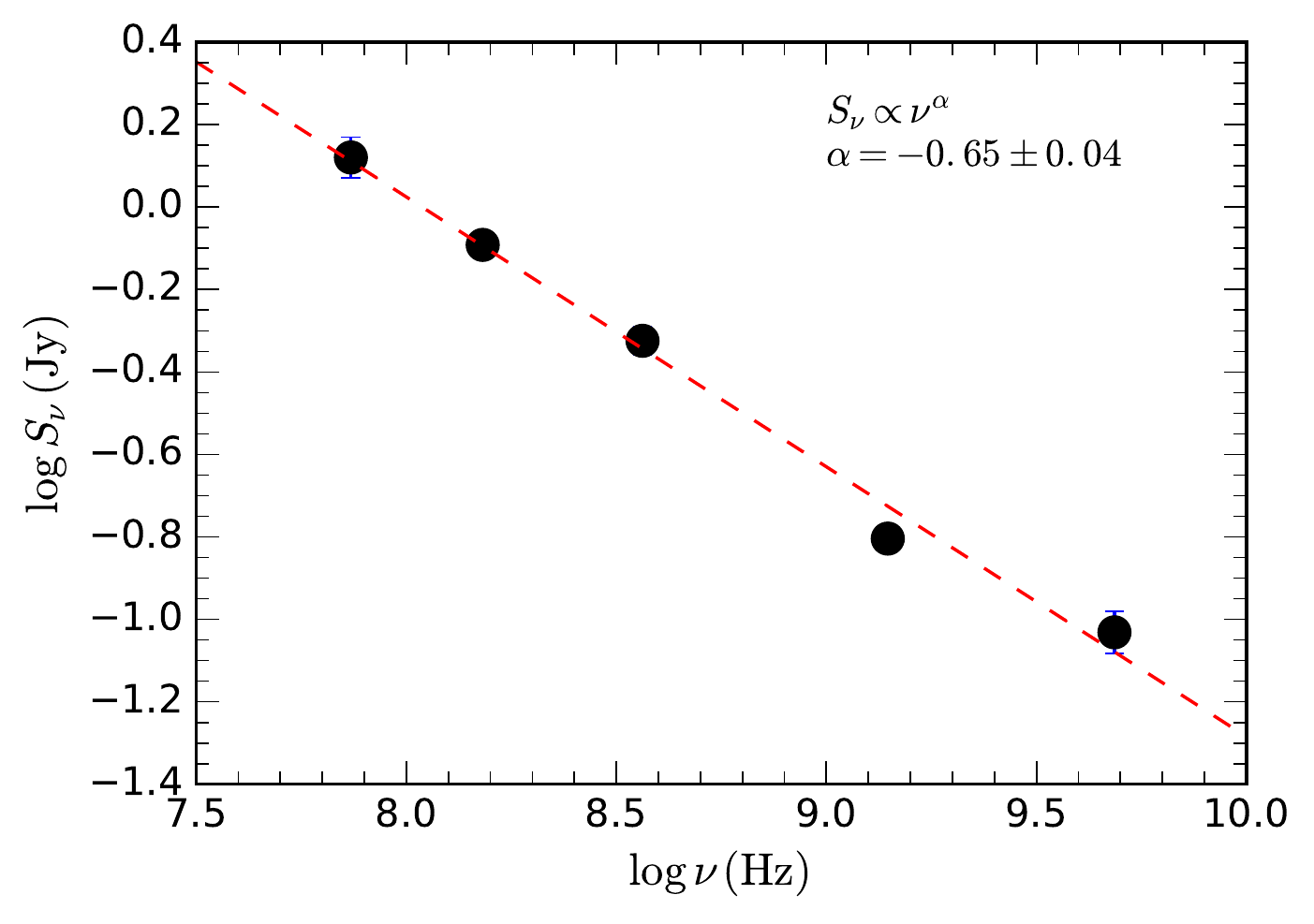}}
\caption{Top: Radio contours from 1.4 GHz FIRST survey have been superposed on SDSS $r$-band image in grey scale. Angular resolution and rms noise in the image are 5.4 arc sec and 0.15 mJy beam$^{-1}$ respectively. Contour levels are plotted as rms noise times ($-$4, $-$2.82, 2.820, 4, 5.650) in steps of $\sqrt{2}$. Bottom:  Radio spectral energy distribution from non-simultaneous data. The best fitted line (S$_{\nu} \propto \boldsymbol{\nu^{\alpha}}$ with 
a power law index of $\alpha=-0.65\pm0.04$ is shown by a dashed line.}\label{Fig:radio_optical_image}. 
\end{figure}

\section{Sample and Data}\label{sec:data}
Our NLSy1 galaxy sample consists of all the 11,101 NLSy1 galaxies cataloged by 
\citet{2017ApJS..229...39R} from SDSS DR 12 spectroscopic data base. 
This is about five fold increase in the number of NLSy1 galaxies known prior to this work from 
\cite{2006ApJS..166..128Z}. To find the radio counterparts  
to the NLSy1 galaxies reported in \cite{2017ApJS..229...39R} we cross matched each NLSy1 galaxy with the 
FIRST  catalog (Catalog version 14dec17\footnote{\url{http://sundog.stsci.edu/first/catalogs.html}}) around a circular region of radius 2 arcsec. This search provided us 
with a sample of 555 NLSy1 galaxies detected by FIRST. We then visually examined each of these 555 FIRST detected NLSy1 galaxies 
for the presence of  extended radio emission. This careful examination led to the discovery of  one source, J103024.95+551622.7, with a 
very large extended radio emission. The core coincides with the optical nucleus with 
extended emission on both sides of the radio core. 

\section{Results}\label{sec:results}
We report the detection of a large kpc scale structure in the RL-NLSy1 galaxy based on the analysis of the FIRST images. It was discovered in the course of our analysis of the radio properties of the new sample of NLSy1 galaxies recently published by
\cite{2017ApJS..229...39R}. 
Figure \ref{Fig:radio_optical_image} shows the radio contours of J103024.95+551622.7
from the FIRST image at 1.4 GHz superposed on its SDSS $r$-band image.

\subsection{Radio Structure}
The source has a radio core ($\alpha_{2000}$ = 10:30:24.787,
$\delta_{2000}$ = 55:16:18.48 with peak flux = 14 mJy beam$^{-1}$) and 
extended radio structures. The angular separation between the 
north-east (NE) (72 mJy beam$^{-1}$) and south-west (SW) (21 mJy beam$^{-1}$) 
peaks is 20.5 arc sec which for the redshift of the host galaxy corresponds to a projected size\footnote{Hereafter, all the sizes are projected size unless specified otherwise.} of 116 kpc. Along with peaks at the lobes and the radio core coinciding with
the host 
galaxy seen in the SDSS $r$-band image, there is significant diffuse emission 
present on either sides of the double lobe. This is a strong evidence that 
the NE and SW peaks are not unrelated point sources but they constitute a 
single radio galaxy with FR II like radio structure with back flow from 
hotspot regions causing the diffuse emission. Integrated flux calculated for 
the whole source from the FIRST image is 155 mJy. So the diffuse emission 
(subtracting the core and peaks at lobes from the integrated flux 
($155-(14+72+21)= 48$ mJy) constitute a major 30\% of the whole radio source 
flux density. It may also be noted that the radio source (lobe-core-lobe) does 
not lie in a straight line and is slightly bent with the NE lobe brighter than the SW counterpart. Also, the presence of diffuse radio emission  at the resolution of FIRST 
can be  established from the dimensionless concentration parameter \citep{2002AJ....124.2364I}, 
$\theta_{\mathrm{FIRST}}=\sqrt{S_{\mathrm{int}}/S_{\mathrm{peak}}}$, where $S_{\mathrm{int}}$ and $S_{\mathrm{peak}}$ are the integrated and peak flux 
densities of the source respectively. $\theta_{\mathrm{FIRST}}$ is found to be 1.69, much larger than 
the value of 1.06 above which a source is defined 
as ``resolved'' by \citet{2008AJ....136..684K}, indicating an extended emission. 
Since, FIRST is a high resolution survey, it is insensitive to the very extended 
emission originating from the lobes. However, NVSS at the same 
frequency provides low resolution images thus detecting much extended emission. 
Therefore, following \citet{2015MNRAS.446..599S}, we estimated 
$\theta_{\mathrm{NVSS-FIRST}}=\sqrt{S_{\mathrm{NVSS,int}}/S_{\mathrm{FIRST, int}}}$, where $S_{\mathrm{NVSS,int}}$ and $S_{\mathrm{FIRST, int}}$ are the integrated flux densities obtained from NVSS and FIRST images respectively. The value of $\theta_{\mathrm{NVSS-FIRST}}$ turns out to be 2.18 suggesting the presence of additional faint 
low-surface-brightness radio component in the source apart from the extended 
radio emission detected in FIRST.

To estimate the radio spectral index ($\alpha$) we collected the 
multi-frequency radio data from the NASA/IPAC Extragalactic 
Database (NED)\footnote{http://nedwww.ipac.caltech.edu} and estimated $\alpha$ using data from 74 MHz to 5 GHz utilizing a linear least-square fit in the log-log 
plane. This is shown in the bottom panel of Figure \ref{Fig:radio_optical_image}. A steep radio spectra ($S_{\nu} \propto \nu^{\alpha}$) 
with $\alpha=-0.65 \pm 0.04$ has been found for this 
object suggesting an extended emission originating from the young radio lobes.

\subsection{Variability}
\begin{figure}
\centering
\resizebox{9.0cm}{4.0cm}{\includegraphics{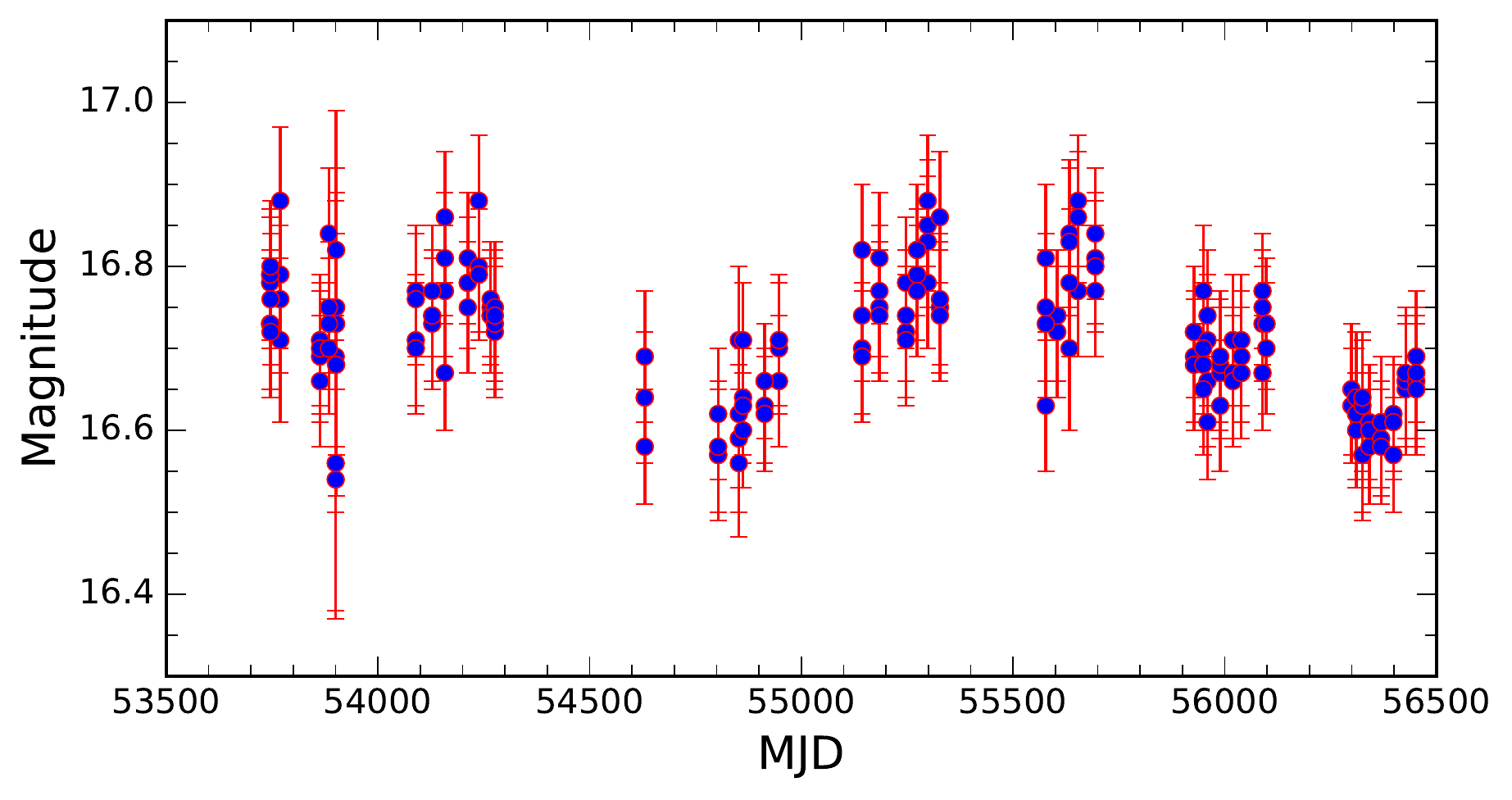}}
\resizebox{8.7cm}{5.5cm}{\includegraphics{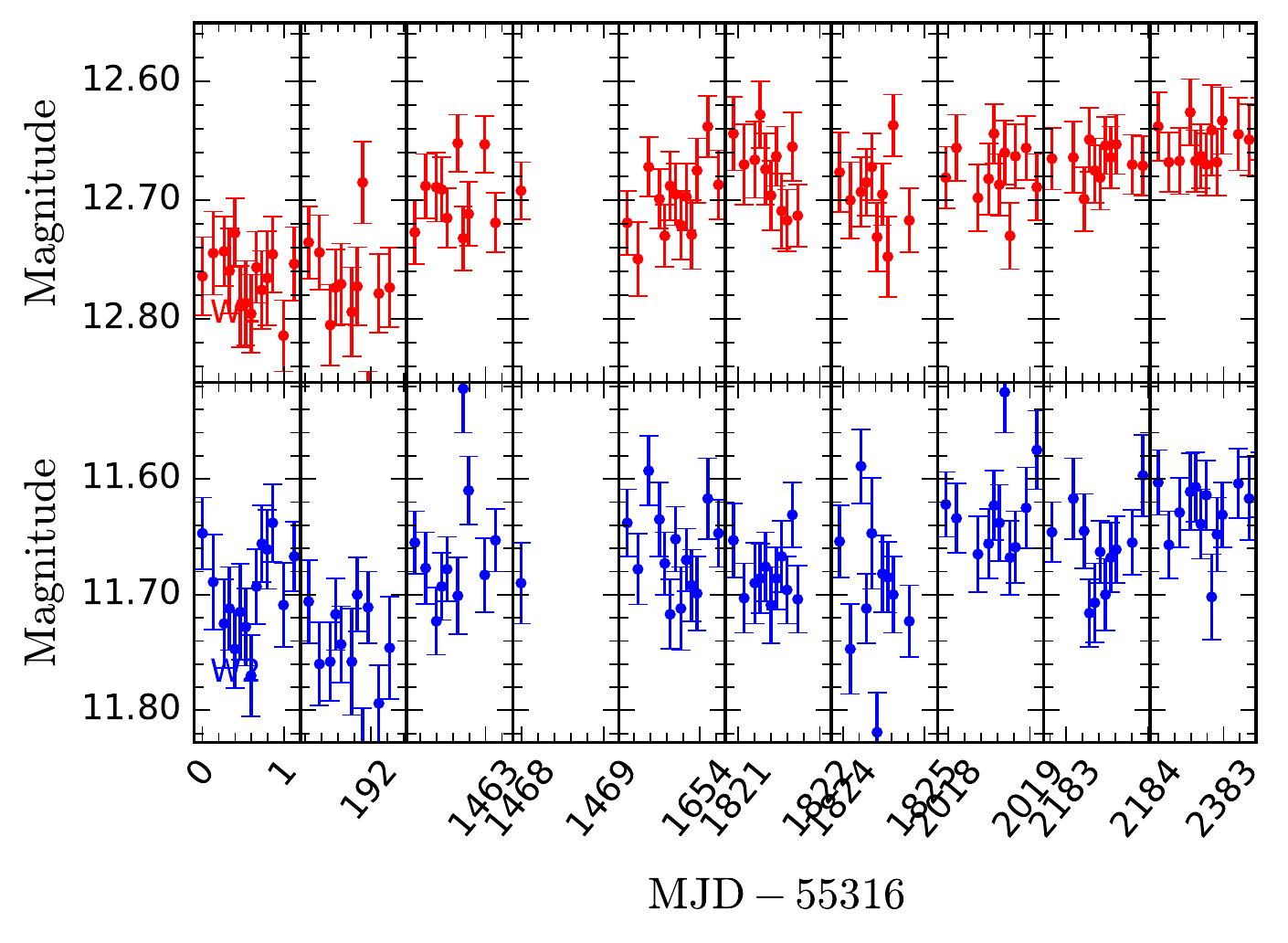}}
\caption{Top: Optical $V$-band light curve of SDSS J103024.95+551622.7 obtained by CRTS. Bottom: Infrared 3.4$\mu$m (upper) and 4.6$\mu$m (lower) light curves.}\label{Fig:lc}. 
\end{figure}

The variability of the source in the optical and mid-IR bands was also
investigated using data from the Catalina Real Time Transient Survey
(CRTS, \citealt{2009ApJ...696..870D} and the 
{\it Wide-field Infrared Survey Explorer} \citep[WISE;][]{2010AJ....140.1868W}
database, respectively. The CRTS optical light curve is 
shown in the top panel of Figure \ref{Fig:lc}. The amplitude of variability 
($\sigma_{\rm m}$) was calculated following \citet{2007AJ....134.2236S} \citep[see also][]{2010ApJ...716L..31A,2017ApJ...842...96R}. 
The source is found to be non-variable in optical. It is also found
to be non-variable in the mid-infrared $W$1 and $W$2 bands both on 
short time scales ($\sim$ 1 day) and long time scales ($\sim$ 7 years).

\begin{figure}
\centering
\resizebox{9cm}{7.0cm}{\includegraphics{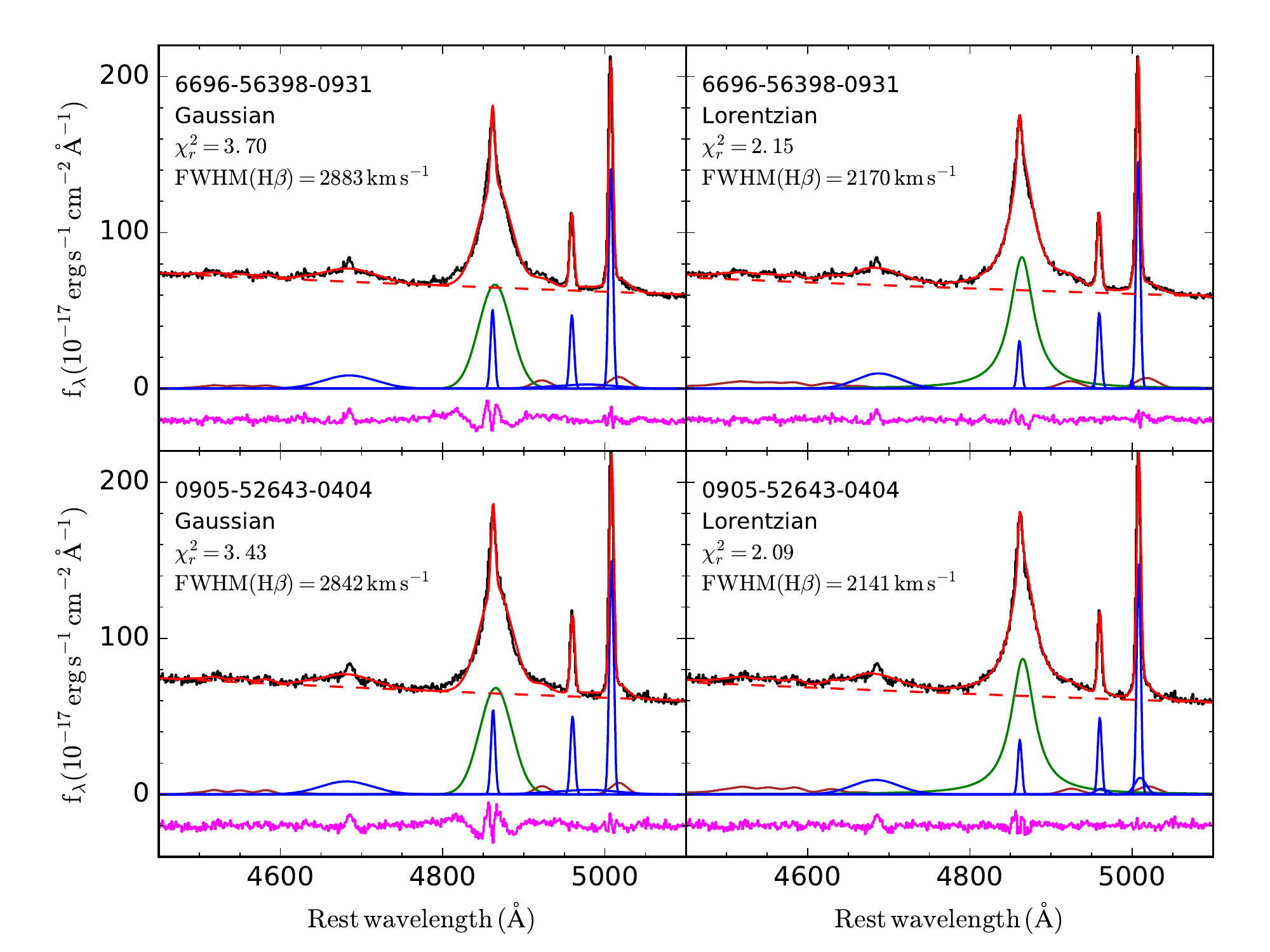}}
\caption{Spectral fittings of SDSS J103024.95+551622.7 around H$\beta$ emission line regions for two different epochs are shown in upper and lower panel with the labels indicating PLATE-MJD-FIBER of SDSS. Broad emission line is fitted using a Gaussian (left) and a Lorentzian (right).  
The observed spectrum (black) and the overall fitted spectrum (red) are shown with the decomposed individual components; broad H$\beta$ is in green, narrow H$\beta$, [O III] doublet and broad He II lines are in blue, Fe II lines are in brown. The power law continuum is shown by red-dashed line. The residual is 
shown by magenta color and the $\chi^2$ of the fit is shown in each panel.}\label{Fig:spec_fit}. 
\end{figure}

\subsection{Optical spectra}
To check on the NLSy1 galaxy classification
of SDSS J103024.95+551622.7, we collected the SDSS spectra, removed the effects
of Galactic extinction, brought it to the rest frame, and re-analysed it using 
the spectral decomposition procedure described by \citet{2017ApJS..229...39R} 
but without subtracting the host galaxy component, which is negligible as the 
redshift of the object is $z=0.4347$. The source was observed twice, hence we 
fitted both the spectra as shown by the black line in the upper and lower panels 
of Figure \ref{Fig:spec_fit} for two different cases; when broad H$\beta$ 
component was fitted using a Gaussian function (left) and a Lorentzian 
function (right). A Lorentzian function provides a better fit to the spectra
having lower reduced-$\chi^2$ value in both the spectra. The use of Lorentzian 
function to fit the broad component of the H$\beta$ emission naturally provides 
narrower FWHM which are $2170\pm 45$ km s$^{-1}$ and $2140\pm 38$ km s$^{-1}$ 
for the two epochs 
compared to $2883\pm 34$ km s$^{-1}$ and $2842\pm 70$ km s$^{-1}$, respectively, found using a 
Gaussian function. The canonical definition of NLSy1 galaxies pertain to 
sources with  FWHM(H$\beta$) $<$ 2000 km s$^{-1}$ \citep{1989ApJ...342..224G}.  However, the criteria 
of FWHM(H$\beta$) $<$ 2200 km s$^{-1}$ has been used in the initial catalog 
of NLSy1 galaxies by \citet{2006ApJS..166..128Z} and recently by 
\cite{2017ApJS..229...39R} since the distribution of broad H$\beta$ line width is smooth with no sharp cutoff between NLSy1 and BLSy1 galaxies. Moreover, a higher cutoff of FWHM(H$\beta$) $< 4000$ km s$^{-1}$ based on quasar main sequence \citep{2000ApJ...536L...5S}, a luminosity-dependent cutoff value for H$\beta$ line width \citep{2000ApJ...543L.111L,2001A&A...372..730V}, or a Eddington ratio cutoff ($L/L_{\mathrm{EDD}}\ge 0.25$) \citep{2007ApJ...654..754N} have also been suggested for NLSy1 galaxies. The source studied here is close to the 
boundary of 2200 km s$^{-1}$ adopted by \cite{2006ApJS..166..128Z} and \cite{2017ApJS..229...39R}.
However, it has low [OIII]/H${\beta}$ ratio and steep soft X-ray spectral index and a Eddington ratio of $1.01$.  
Putting all the other characteristics together, it is most likely that 
SDSS J103024.95+551622.7 is a NLSy1 galaxy.

\begin{figure}
\centering
\includegraphics[width=1\linewidth]{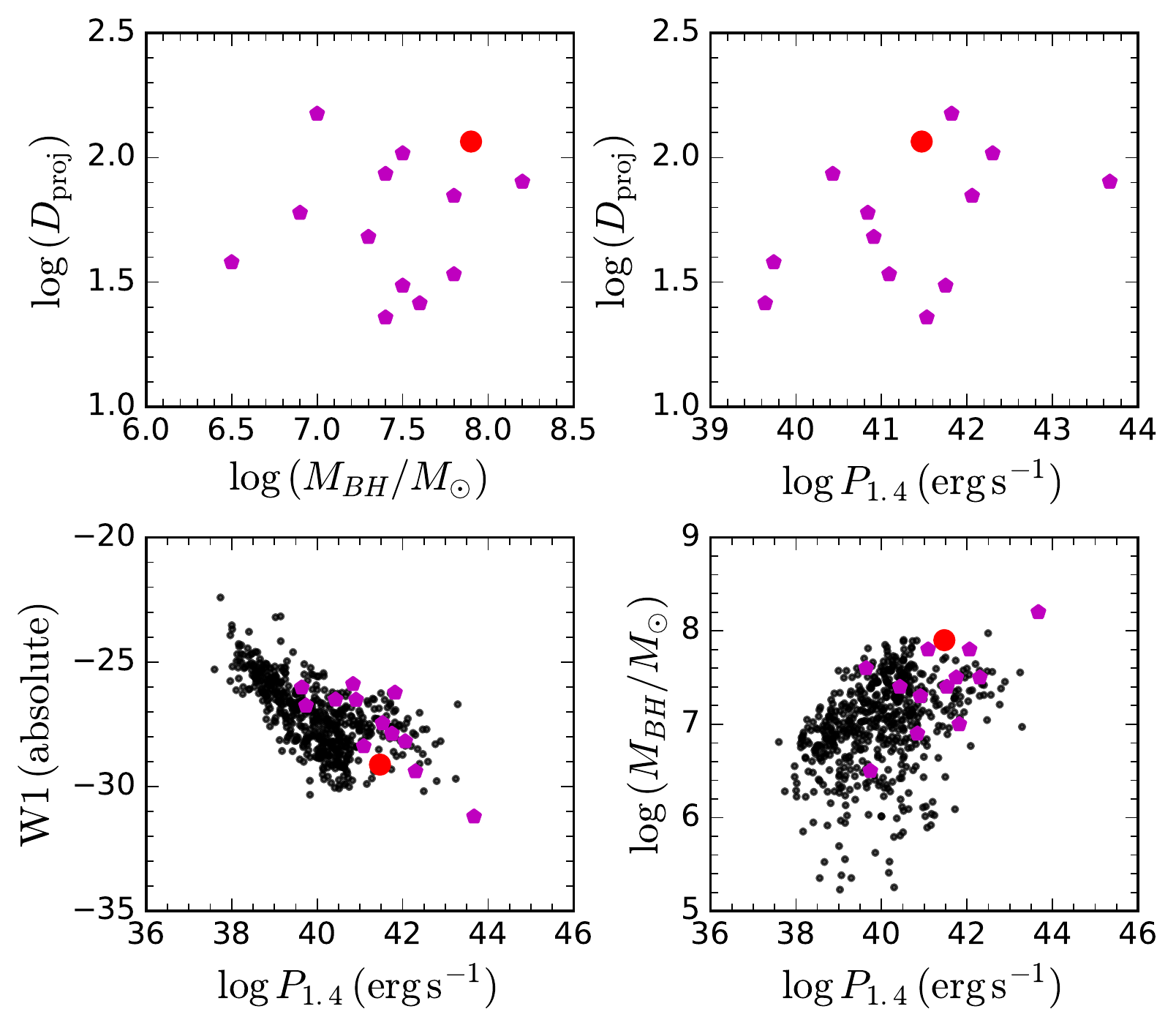}
\caption{Projected size versus black hole mass (upper left) and radio power at 1.4 GHz (upper right). WISE $W$1-band magnitude versus radio power is shown in 
lower left, and the black hole mass versus radio power is shown in the lower 
right. The red large symbol represents SDSS J103024.95+551622.7. The polygons represent NLSy1 galaxies having larger than 20 kpc radio structure (see Table \ref{table:obj}) except PKS 1502+036. The points in the lower panels represent NLSy1 galaxies detected in FIRST from \citet{2017ApJS..229...39R}. In the lower left panel, only 
objects having S/N ratio $>$10 in WISE $W$1-band are plotted.}
\label{fig:plot_corr}
\end{figure}

\section{Interpretation and discussion}\label{sec:discussion}

\subsection{Comparison with a sample of NLSy1 galaxies}

We collected various information of all the NLSy1 galaxies having larger than 20 kpc radio 
emission from the literature and tabulated them in Table \ref{table:obj}. Note 
that one of the sources, NVSS J095317+283601  
in Table \ref{table:obj} has FHWM of H$\beta$ (2162 $\pm$ 201 km s$^{-1}$)
similar to the value of SDSS J103024.95+551622.7. A higher FWHM (H$\beta$) will lead to a higher black hole mass. However, all the sources except SDSS J122222.55+041315.7 in Table \ref{table:obj} have black hole masses $<10^8 \, M_{\odot}$.  We 
investigated the relationship between projected size of the radio jet 
($D_{\mathrm{proj}}$) with $M_{\mathrm{BH}}$ and radio power at 1.4 GHz ($P_{1.4}$) for NLSy1 
galaxies in 
Figure \ref{fig:plot_corr}. The red point in each panel represents 
SDSS J103024.95+551622.7. The magenta polygons represent NLSy1 galaxies 
that have larger than 20 kpc extended radio structures 
(see Table \ref{table:obj}). No correlation was found between the
projected size both with $M_{\mathrm{BH}}$ (or jet power). 
When the sources are put in the WISE $W$1-band absolute magnitude vs. radio power 
diagram, a strong correlation is found. The points in this plot are the 
NLSy1 galaxies from \citet{2017ApJS..229...39R} which are detected in FIRST 
and WISE in $W$1-band with S/N $>$10. The Spearman correlation test on the full 
sample yields a correlation coefficient ($r$) of $-0.6$ and a very low $p-$ 
value ($p<1\times10^{-20}$) suggesting a significant negative correlation 
where absolute IR magnitude decreases with radio power. The black hole mass of 
all those objects is plotted against their radio power on the right panel. We 
found a positive correlation having $r=0.4$ which is similar to the correlation
found by \citet{2006ApJ...637..669L} between jet power and black hole mass 
suggesting high powered jets originate from massive AGN. From Figure \ref{fig:plot_corr} (bottom right panel) 
it is evident that NLSy1 galaxies with $>$ 20 kpc radio structure predominantly 
occupy the region with large radio power and high black hole mass. In 
terms of the projected radio size SDSS J103024.95+551622.7 is the second 
largest NLSy1 galaxy after SDSS J110006.07+442144.3\footnote{Note that 
SDSS J110006.07+442144.3 does not satisfy $\mathrm{[O III]/H\beta}$ flux ratio 
$< 3$ \citep{2014ApJ...793L..26T} required for NLSy1 galaxy classification.}. In Figure \ref{fig:plot_corr}, it follows the $W1-P_{1.4}$ 
anti-correlations and $M_{\mathrm{BH}}-P_{1.4}$ correlation, thus not 
different from other kpc scale radio emitting NLSy1 galaxies.

\subsection{A standard powerful radio galaxy}

From the radio spectrum in Figure \ref{Fig:radio_optical_image}, we see that the spectrum is steep. Considering the lobes of radio galaxies which 
are optically thin, the derived spectral index of $-$0.65 $\pm$ 0.04 is very close to the theoretical 
value of injection spectral index of around $-$0.62\footnote{A negative sign is used to match the definition of spectral index.} 
\citep{2000ApJ...542..235K,2013MNRAS.430.2137K}. If the radio jets have the same spectral index of $-$0.65 then the inverse-Compton effect of
the cosmic microwave background (CMB) and radio photons together can give rise 
to a X-ray power-law
spectrum in the 0.1 to 10 kev range whose photon index is supposed 
to be $-$(1+0.65) = $-$1.65 \citep{2009MNRAS.400..480K}. However, for
the source SDSS J103024.95+551622.7 the photon index estimated from ROSAT data is $-$2.61 which is 
much steeper. To resolve this inconsistency, we need further observations
in the 0.1 to 10 keV band. The presence of strong [O III] doublet, as
seen in Figure \ref{Fig:spec_fit}, indicates strong black body emission from the accretion 
disk, suggesting a standard accretion disk through which the matter is 
being accreted onto the black hole of this source. Therefore, this
NLSy1 radio galaxy is a High Excitation Radio Galaxy (HERG) in which 
cold mode accretion is taking place \citep{2006MNRAS.372...21A}. 
Assuming the radio spectral index
($-$0.65) of this radio galaxy to be the injection spectral 
index ($\alpha_{\mathrm{inj}}$),
the correlation between $\alpha_{\mathrm{inj}}$ and jet power $Q_{\mathrm{jet}}$, as 
published by \cite{2013MNRAS.436.1595K}, in their Figure 2, we estimated
a jet power of this source to be $3\times 10^{44}$ erg s$^{-1}$. This 
suggests that this NLSy1 galaxy has a jet power typical of classical
radio galaxies.

In the lower left panel of Figure \ref{fig:plot_corr}, we see that there is a strong correlation
between $W$1 and $P_{1.4}$. Most of the luminosity in the $W$1-band of 
WISE comes from the old stars in the bulge. This means, more the $W$1
luminosity, heavier is the bulge. The heavier the bulge heavier the black
hole mass \citep{1995ARA&A..33..581K}. 
Heavier the black hole mass, we might get higher jet power provided the Eddington scaled mass 
accretion rate and the spin of the black holes are in a narrow range.
We indeed get a correlation between black hole mass and the radio power
at 1.4 GHz (which is a proxy of jet power).
Therefore, we can expect that all NLSy1 galaxies are accreting with
Eddington scaled mass accretion rate within a very narrow range. If spin 
of the black hole has a role in deciding the jet power, then all
the NLSy1 galaxies must have spin within a very narrow range (or jet power 
has no dependence on the spin). Therefore, the correlations in the left 
and right bottom panels are quite expected. 

Similarly, the absence of correlation in the top right and top left panels 
are also quite expected. Let us consider the top left panel, where source 
size vs. $M_{\mathrm{BH}}$ has been plotted.  If the source size would have 
depended only on the jet power, then we would have perhaps got a correlation
between source size and the black hole mass. However, the source size 
depends on jet power ($Q_{\mathrm{jet}}$), density ($\rho_{\mathrm{amb}}$) of ambient medium 
and the angle ($\theta$) of the jet with the Line of Sight (LoS). Even if
the jet power is correlated with the black hole mass, the scatter of 
the values of $\rho_{\mathrm{amb}}$ and $\theta$ from source to source will spoil
any correlation between source size and the black hole mass. Since 
black hole mass and $P_{1.4}$ are correlated, it is not expected to get 
any correlation between source size and $P_{1.4}$, as there is no 
correlation between source size and black hole mass. A larger sample will help to address this.

\begin{figure}
\centering
\includegraphics[width=1\linewidth]{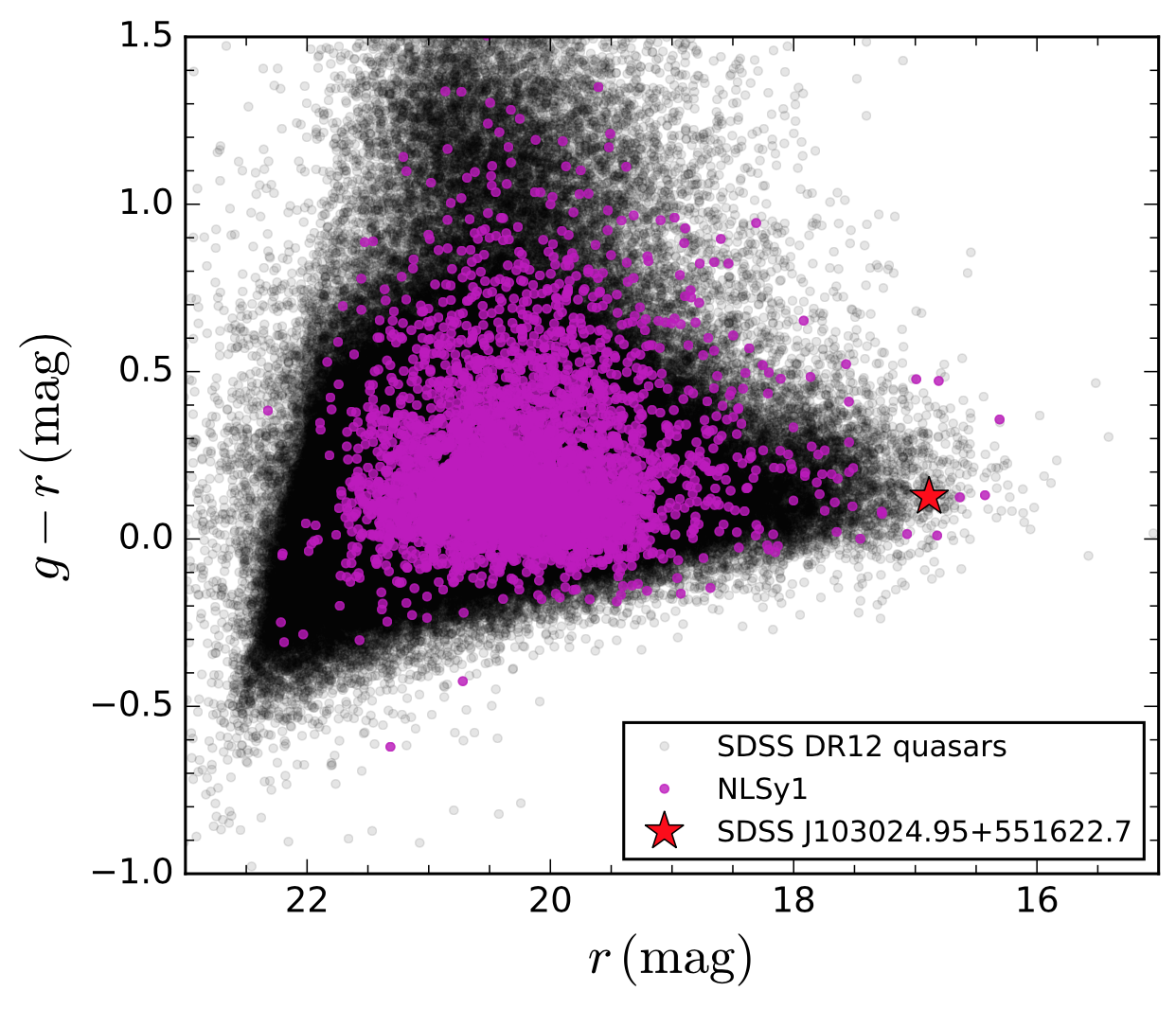}
\caption{Optical $g-r$ color versus $r$ magnitude diagram. The NLSy1 galaxies (magenta dots) along with SDSS quasars (black dots) and SDSS J103024.95+551622.7 (star) is shown.}
\label{fig:plot_color_mag}
\end{figure}

\subsection{Link with Speca:} NLSy1 galaxies are more often associated with 
spiral or disk galaxies with rich gaseous central regions
\citep{2006AJ....132..321D, 2007ApJS..169....1O}. Not only in  NLSy1 galaxies but also in general Seyfert galaxies, 
radio-loud galaxies with over 100 kpc large radio lobes are rare. 
While almost all large (over 100 kpc) radio galaxies are hosted in elliptical 
galaxies only {\it six}  
of them are known to be hosted in disk/spiral galaxies with large radio lobes. In Figure \ref{fig:plot_color_mag}, we plotted the optical $g-r$ color against $r$ magnitude of NLSy1 galaxies from \citet{2017ApJS..229...39R} in dots (magenta) along with SDSS quasars in the background (black dots). Our target is represented by a star symbol having a $g-r$ color of 0.10 mag. According to \citet{2010MNRAS.406..342B}, a galaxy can be defined as a spiral if the observer-frame $g - r < 0.6$ mag \citep[see also][]{2011A&A...529A..53T}. Thus, SDSS J103024.95+551622.7 can be considered as a spiral galaxy following the above criteria. 
According to the plot, majority of the NLSy1 galaxy is expected to be spiral. 
However, deep imaging observations of NLSy1 galaxies are needed to confirm if the hosts of NLSy1 
galaxies are indeed spirals. Nevertheless, a few examples of spiral host radio galaxies are known such as J0313-192 \citep{1998ApJ...495..227L}, 
Speca \citep{2011MNRAS.417L..36H}, J0314-1906, J2345-0449, J0836+0532 and MCG+07-47-10 \citep{2016JApA...37...41H}. It is likely that such spiral 
host radio galaxies could be more frequent in the early Universe, 
prior to the quasar/radio galaxy era ($z$ $\sim$ 2$-$3 ) and when spiral galaxies were more prevalent than the merger-remnant 
ellipticals \citep{2016JApA...37...41H}. Thus, this new object reported here 
SDSS J103024.95+551622.7, and hosted in a spiral galaxy may represent an opportunity to investigate non-standard modes of 
relativistic jet production in passively evolving galaxies with smaller black holes masses 
which may have been more common in the early Universe, prior to the rise of merger-remnant ellipticals and supermassive black holes. 

\subsection{Doppler factor, inclination angle and age}
The observed core dominance parameter ($r_{obs}$; \citealt{1982MNRAS.200.1067O}) can be calculated from the ratio of core (S$_{\mathrm{core}}$) to lobe 
(S$_{\mathrm{lobe}}$) flux densities. At 1.4 GHz from FIRST, the source has an 
integrated core flux density of 32.7 mJy with the SW and NE lobes having
integrated flux densities of 21.9 mJy and 95.3 mJy respectively. The observed 
core dominance parameter is thus $r_{\mathrm{obs}}=0.28$. This value is higher than that known for radio galaxies
with FR I and FR II radio morphologies having median values $r_{\mathrm{obs}}$ 
= 0.022 and 0.003 respectively \citep{1997MNRAS.284..541M}. We note that 
blazars show extreme core dominance, which is caused by Doppler boosting of 
the core \citep{2015MNRAS.451.4193C}. We calculated Doppler factor ($\delta$) 
following \citet{2012ApJ...760...41D} 
\begin{equation}\label{eq:doppler}
{\delta_{\mathrm{core}}}^{(3-\alpha)} = \frac{r_{\mathrm{obs}}}{r_{\mathrm{int}}},
\end{equation}
where $r_{\mathrm{int}}$ is the intrinsic core dominance parameter and $\alpha$ is the 
core spectral index. We assumed the core spectral index to be 
flat with $\alpha=0$. Calculation of $r_{\mathrm{int}}$ needs an estimation of
the intrinsic core power. We derived the intrinsic core power using the
empirical relation known for radio galaxies and given by 
$\log P_c = (0.62 \pm 0.04) \times \log P_t + (7.6 \pm 1.1)$ 
\citep{2001ApJ...552..508G}, where $P_c$ is the core radio power at 5 GHz and 
$P_t$ is the total radio power at 408 MHz. Using the core flux at 5 GHz and
the total flux at 365 MHz from NED, we found $r_{\mathrm{int}}$  = 0.008. 
This is much lower than $r_{\mathrm{obs}}$ suggesting strong Doppler boosting in 
the core. Using equation \ref{eq:doppler} we estimated  $\delta=3.3$.

To constrain the jet speed ($\beta_{\mathrm{core}}=v_{\mathrm{core}}/c$) and inclination 
angle ($\theta_{\mathrm{core}}$) we used the relation  
$\delta_{\mathrm{core}}= \sqrt{(1-{\beta^2}_{\mathrm{core}})}/(1-\beta_{\mathrm{core}} \,  \cos\theta_{\mathrm{core}})$.   
This relation is highly degenerate, however, we obtained
values of $\beta_{\mathrm{core}} >0.87$ and $\theta_{\mathrm{core}}< 12 \degree$. These values 
are similar to that obtained by \citet{2012ApJ...760...41D} for a sample
of NLSy1 galaxies suggesting that radio-loud NLSy1 galaxies are viewed 
pole-on. Similar low inclination values in the range of $10-15\degree$ were 
also obtained by \citet{2015ApJ...800L...8R} for other kpc-scale jetted 
NLSy1 galaxies. Considering a viewing angle of 12$\degree$, we found the 
deprojected size of the source as $D_{\mathrm{de-proj}} > 116/\sin{12\degree} 
>557$ kpc.

The source exhibits a two-sided structure at kpc scale as seen in 
Figure \ref{Fig:radio_optical_image}. Therefore, the flux ratio of the 
approaching and receding lobes ($R_F$) were used to calculate the speed of the 
kpc scale lobes following \cite{2015ApJ...800L...8R} 
\begin{equation}
R_{F} = \left[\frac{(1+\beta_{\mathrm{kpc}} \cos \theta)}{(1-\beta_{\mathrm{kpc}} \cos \theta)} \right]^{3-\alpha},
\end{equation}
where the spectral index $\alpha=-0.65$. Using, $R_F=4.35$, we 
obtained $\beta_{\mathrm{kpc}} \cos \theta=0.2$. Assuming the inclination angle 
to remain constant between the core and kpc scale lobes, i.e. $\theta=12\degree$, we obtained $\beta_{\mathrm{kpc}} \sim 0.2$. Considering the expansion speed of 
$\beta_{kpc}$ and a two-sided expansion, we estimated the kinetic age of the source 
using $t_{\mathrm{kinetic}} = D_{\mathrm{de-proj}}/2\beta_{\mathrm{kpc}} c$, where $D_{\mathrm{de-proj}}$ is the 
deprojected size of the source. We obtained $t_{\mathrm{kinetic}}$  $>4\times10^6$ yr.

We also estimated the limiting radiative age of this radio source as  
$\tau_{\mathrm{rad}}=50.3 \frac{B^{1/2}}{B^2+B_{\mathrm{IC}}^2} \frac{1}{\sqrt{\nu_{\mathrm{br}}(1+z)} }$, 
where $B$ is the lobe magnetic field expressed in nT,  $B_{\mathrm{IC}}=0.318(1+z)^2$ nT 
is the Cosmic Microwave Background equivalent magnetic field \citep{2006MNRAS.372..693K} 
and  $\nu_{\mathrm{br}}$ is the break frequency (in GHz) observed in the radio spectrum. 
From the above expression, we get the highest radiative age, 
when $B=\frac{B_{\mathrm{IC}}}{\sqrt{3}}$. Using $B=\frac{B_{\mathrm{IC}}}{\sqrt{3}}=0.378$ nT, we 
obtained a radiative age of 20.2 Myr for the break frequency of 5 GHz. Since there 
is no break observed in the radio spectrum, the highest observed radio frequency 
is the lower limit of the break frequency. Thus, $\nu_{\mathrm{br}} > 5$ GHz, 
corresponds to $\tau_{\mathrm{rad}} < 20.2$ Myr. Both the kinetic and 
radiative age measurements agree well and provide a strong constraint on the 
age of SDSS J103024.95+551622.7 to be $> 10^6$ yr. This also
lies within the age range of $10^5 - 10^7$ yr found for other radio-loud
NLSy1 galaxies by \cite{2012ApJ...760...41D} and \citet{2015ApJ...800L...8R}. 
It is thus likely that NLSy1 galaxies are young AGN \citep{2001NewA....6..321M}.

\section{Conclusion}\label{sec:conclusion}
The object SDSS J103024.95+551622.7 has recently been classified as a 
NLSy1 galaxy by \citet{2017ApJS..229...39R} after carefully modeling the spectrum from SDSS DR 12 database. While studying the 1.4 GHz radio images obtained by the FIRST survey, the source is found to have a 116 kpc (projected size) large double lobe FR II like radio structure and 
a central core coinciding with the NLSy1 optical galaxy. The core and diffuse emission, associated with the back flow from lobes, conclusively proves the radio lobes and host galaxy association.  

The source exhibits a steep radio spectrum with $\alpha=-0.65 \pm 0.04$ estimated from the multi-frequency radio data collected from 
NED database.  The source is non-variable in the optical and
mid-infrared bands. 
We modeled the emission spectra of the source obtained by SDSS during its 2 epochs of observation and found that a Lorentzian function can best represent the broad component of H$\beta$ line rather than a Gaussian function. The results obtained from the fitting of two spectra are consistent having width less than 2200 km s$^{-1}$ and F([O III])/ F(H$\beta$) $\sim 0.2$ making it a NLSy1 galaxy as also cataloged by \citet{2017ApJS..229...39R}. 
Though nearly a dozen NLSy1 galaxies have been found to have 20$-$150 kpc extended radio 
emission, {\it none} of them are observed to be fully blown
double-lobed, larger than 100 kpc sources to be compared to 
classical radio galaxies. This is a large (116 kpc) FR II like double 
lobe radio source associated with a NLSy1 as the host galaxy. 
As radio galaxies are typically hosted in ellipticals and extremely rarely 
(only six) with spirals,  understanding of the host or parent population of 
these NLSy1 galaxies with jets and lobes are far from clear. Radio jets are typically 
produced in massive spinning black holes with low accretion rate and hence 
this counter-trend NLSy1 galaxy, SDSS J103024.95+551622.7 with larger than 
100 kpc radio lobes may help to understand the unexplored parameter space of jet 
production mechanism and AGN feedback-driven galaxy evolution.

\begin{center}
\begin{table*}
\resizebox{1.0\linewidth}{!}{%
\begin{threeparttable}
\caption {Details of radio-loud NLSy1 galaxies with larger than 20 kpc scale radio structures. Those marked with * are sources that are emitters of $\gamma$-rays. The columns are (1) object name; (2) Right ascension; (3) Declination; (4) redshift; (5) H$\beta$ line FWHM (km s$^{-1}$); (6) logarithmic black hole mass ($M_{\odot}$); (7) Projected size of the radio structure (kpc); (8) logarithmic radio power at 1.4 GHz (erg s$^{-1}$); (9) $\theta_{\mathrm{FIRST}}$; (10) radio spectral index; (11) WISE magnitude in $W$1-band (2.4 $\mu m$) and (12) References.}
\begin{tabular}{lccccccccccc} \hline
Name                          & RA (2000) & Dec (2000) & z     &  FWHM   & log($M_{BH}/M_{\odot}$) & $D_{\mathrm{proj}}$ & $\log P_{1.4}$ & $\theta_{\mathrm{FIRST}}$ &  $\alpha$ &  W1  & Ref. \\
      &       &     &    &  (km s$^{-1}$) &   &  (kpc) & (erg s$^{-1}$) &       &              & (mag)  & \\ 
(1)   & (2)   & (3) & (4) & (5) & (6) & (7) & (8)  & (9)  & (10) & (11) & (12) \\ \hline
1H 0323+342*                  & 03:24:41.1 & $+$34:10:46 & 0.063 & 1520 & 7.3   & 48    & 40.91 & 1.21  & 0.1      & 10.74 $\pm$ 0.02 &  \citet{2012ApJ...760...41D}  \\
PKS 0558-504                  & 05:59:47.4 & $-$50:26:52 & 0.137 & 1250 & 7.8   & 34    & 41.09 & 1.18  & $-$0.3   & 10.67 $\pm$ 0.02 &  \citet{2010ApJ...717.1243G} \\
PMN J0948+0022*               & 09:48:57.3 & $+$00:22:26 & 0.585 & 1432 & 7.5   & 104   & 42.30 & 1.01  & 0.77     & 13.28 $\pm$ 0.02 &  \citet{2012ApJ...760...41D} \\
NVSS J095317+283601           & 09:53:17.1 & $+$28:36:02 & 0.659 & 2162 & 7.8   & 70.2  & 42.06 & 1.03  & $-$0.50  & 14.80 $\pm$ 0.03 &  \citet{2015ApJ...800L...8R},  \\
\it{SDSS J103024.95+551622.7} & 10:30:24.9 & $+$55:16:23 & 0.435 & 2170 & 7.9   & 116.0 & 41.47 & 1.69  & $-$0.65  & 12.78 $\pm$ 0.02 &  This work \\
SDSS J110006.07+442144.3      & 11:00:06.1 & $+$44:21:44 & 0.84  & 1900 & 7.0   & 150   & 41.82 & 1.37  & 0.20     & 17.40 $\pm$ 0.13 &  \citet{2014ApJ...793L..26T}, \citet{2017arXiv170907202G} \\
SDSS J120014.08-004638.7      & 12:00:14.1 & $-$00:46:39 & 0.179 & 1945 & 7.4   & 86    & 40.43 & 1.81  & ---      & 13.18 $\pm$ 0.02 &  \citet{2012ApJ...760...41D} \\
SDSS J122222.55+041315.7*     & 12:22:22.5 & $+$04:13:16 & 0.966 & 1734 & 8.2   & 80    & 43.67 & 1.03  & 0.30     & 12.80 $\pm$ 0.02 &  \citet{2015MNRAS.454L..16Y}  \\
Mrk 783                       & 13:02:58.8 & $+$16:24:27 & 0.067 & 770  & 7.6   & 26.0  & 39.64 & 1.24  & 0.67     & 11.37 $\pm$ 0.02 &  \citet{2017arXiv170403881C} \\
NVSS J143509+313149           & 14:35:09.5 & $+$31:31:48 & 0.502 & 1719 & 7.5   & 30.6  & 41.75 & 1.06  & $-$0.72  & 14.39 $\pm$ 0.03 &  \citet{2015ApJ...800L...8R} \\
SDSS J145041.93+591936.9      & 14:50:41.9 & $+$59:19:37 & 0.202 & 1159 & 6.5   & 38    & 39.74 & 1.02  & ---      & 13.23 $\pm$ 0.02 &  \citet{2012ApJ...760...41D} \\
PKS 1502+036*                 & 15:05:06.5 & $+$03:26:31 & 0.409 & 1082 & 6.6   & $<$25 & 42.46 & 1.02  & 0.70     & 14.02 $\pm$ 0.02 &  \cite{2013MNRAS.433..952D} \\
FBQS J1644+2619*              & 16:44:42.5 & $+$26:19:13 & 0.144 & 1507 & 6.9   & 60    & 40.84 & 1.01  & 0.38     & 13.28 $\pm$ 0.02 &  \citet{2012ApJ...760...41D}  \\
NVSS J172206+565452           & 17:22:06.0 & $+$56:54:52 & 0.426 & 1385 & 7.4   & 22.8  & 41.53 & 1.03  & $-$0.64  & 14.39 $\pm$ 0.03 &  \citet{2015ApJ...800L...8R} \\
\hline
\end{tabular}
\end{threeparttable}}
\label{table:obj}
\end{table*}
\end{center}
      

\acknowledgments
We thank the anonymous referee for his/her critical comments that helped
to improve the manuscript. S.R. acknowledges the support by the Basic Science Research Program through the National Research Foundation of Korea government (2016R1A2B3011457). S.R. thanks Neha Sharma (KHU, South Korea) for carefully reading the manuscript.
\bibliographystyle{apj}
\bibliography{ref}

\begin{thebibliography}{}
\expandafter\ifx\csname natexlab\endcsname\relax\def\natexlab#1{#1}\fi

\bibitem[{{Abdo} {et~al.}(2009){Abdo}, {Ackermann}, {Ajello}, {Axelsson},
  {Baldini}, {Ballet}, {Barbiellini}, {Bastieri}, {Battelino}, {Baughman},
  {Bechtol}, {Bellazzini}, {Bloom}, {Bonamente}, {Borgland}, {Bregeon}, {Brez},
  {Brigida}, {Bruel}, {Caliandro}, {Cameron}, {Caraveo}, {Casandjian},
  {Cavazzuti}, {Cecchi}, {Chekhtman}, {Cheung}, {Chiang}, {Ciprini}, {Claus},
  {Cohen-Tanugi}, {Collmar}, {Conrad}, {Costamante}, {Dermer}, {de Angelis},
  {de Palma}, {Digel}, {Silva}, {Drell}, {Dubois}, {Dumora}, {Farnier},
  {Favuzzi}, {Focke}, {Foschini}, {Frailis}, {Fuhrmann}, {Fukazawa}, {Funk},
  {Fusco}, {Gargano}, {Gehrels}, {Germani}, {Giebels}, {Giglietto}, {Giordano},
  {Giroletti}, {Glanzman}, {Grenier}, {Grondin}, {Grove}, {Guillemot},
  {Guiriec}, {Hanabata}, {Harding}, {Hartman}, {Hayashida}, {Hays}, {Hughes},
  {J{\'o}hannesson}, {Johnson}, {Johnson}, {Johnson}, {Kamae}, {Katagiri},
  {Kataoka}, {Kerr}, {Kn{\"o}dlseder}, {Kuehn}, {Kuss}, {Lande}, {Latronico},
  {Lemoine-Goumard}, {Longo}, {Loparco}, {Lott}, {Lovellette}, {Lubrano},
  {Madejski}, {Makeev}, {Max-Moerbeck}, {Mazziotta}, {McConville}, {McEnery},
  {Meurer}, {Michelson}, {Mitthumsiri}, {Mizuno}, {Monte}, {Monzani},
  {Morselli}, {Moskalenko}, {Murgia}, {Nolan}, {Norris}, {Nuss}, {Ohsugi},
  {Omodei}, {Orlando}, {Ormes}, {Paneque}, {Panetta}, {Parent}, {Pavlidou},
  {Pearson}, {Pepe}, {Pesce-Rollins}, {Piron}, {Porter}, {Rain{\`o}}, {Rando},
  {Razzano}, {Readhead}, {Reimer}, {Reimer}, {Reposeur}, {Richards}, {Ritz},
  {Rodriguez}, {Romani}, {Ryde}, {Sadrozinski}, {Sambruna}, {Sanchez},
  {Sander}, {Parkinson}, {Scargle}, {Schalk}, {Sgr{\`o}}, {Smith}, {Spandre},
  {Spinelli}, {Starck}, {Stevenson}, {Strickman}, {Suson}, {Tagliaferri},
  {Takahashi}, {Tanaka}, {Thayer}, {Thompson}, {Tibaldo}, {Tibolla}, {Torres},
  {Tosti}, {Tramacere}, {Uchiyama}, {Usher}, {Vilchez}, {Vitale}, {Waite},
  {Winer}, {Wood}, {Ylinen}, {Zensus}, {Ziegler}, {Fermi/LAT Collaboration},
  {Ghisellini}, {Maraschi}, {Tavecchio}, \& {Angelakis}}]{2009ApJ...699..976A}
{Abdo}, A.~A., {Ackermann}, M., {Ajello}, M., {et~al.} 2009, \apj, 699, 976

\bibitem[{{Ai} {et~al.}(2010){Ai}, {Yuan}, {Zhou}, {Wang}, {Dong}, {Wang}, \&
  {Lu}}]{2010ApJ...716L..31A}
{Ai}, Y.~L., {Yuan}, W., {Zhou}, H.~Y., {et~al.} 2010, \apjl, 716, L31

\bibitem[{{Alam} {et~al.}(2015){Alam}, {Albareti}, {Allende Prieto}, {Anders},
  {Anderson}, {Anderton}, {Andrews}, {Armengaud}, {Aubourg}, {Bailey}, \&
  et~al.}]{2015ApJS..219...12A}
{Alam}, S., {Albareti}, F.~D., {Allende Prieto}, C., {et~al.} 2015, \apjs, 219,
  12

\bibitem[{{Allen} {et~al.}(2006){Allen}, {Dunn}, {Fabian}, {Taylor}, \&
  {Reynolds}}]{2006MNRAS.372...21A}
{Allen}, S.~W., {Dunn}, R.~J.~H., {Fabian}, A.~C., {Taylor}, G.~B., \&
  {Reynolds}, C.~S. 2006, \mnras, 372, 21

\bibitem[{{Ant{\'o}n} {et~al.}(2008){Ant{\'o}n}, {Browne}, \&
  {March{\~a}}}]{2008A&A...490..583A}
{Ant{\'o}n}, S., {Browne}, I.~W.~A., \& {March{\~a}}, M.~J. 2008, \aap, 490,
  583

\bibitem[{{Baldi} {et~al.}(2016){Baldi}, {Capetti}, {Robinson}, {Laor}, \&
  {Behar}}]{2016MNRAS.458L..69B}
{Baldi}, R.~D., {Capetti}, A., {Robinson}, A., {Laor}, A., \& {Behar}, E. 2016,
  \mnras, 458, L69

\bibitem[{{Banerji} {et~al.}(2010){Banerji}, {Lahav}, {Lintott}, {Abdalla},
  {Schawinski}, {Bamford}, {Andreescu}, {Murray}, {Raddick}, {Slosar},
  {Szalay}, {Thomas}, \& {Vandenberg}}]{2010MNRAS.406..342B}
{Banerji}, M., {Lahav}, O., {Lintott}, C.~J., {et~al.} 2010, \mnras, 406, 342

\bibitem[{{Berton} {et~al.}(2018){Berton}, {Congiu}, {J{\"a}rvel{\"a}},
  {Antonucci}, {Kharb}, {Lister}, {Tarchi}, {Caccianiga}, {Chen}, {Foschini},
  {L{\"a}hteenm{\"a}ki}, {Richards}, {Ciroi}, {Cracco}, {Frezzato}, {La Mura},
  \& {Rafanelli}}]{2018arXiv180103519B}
{Berton}, M., {Congiu}, E., {J{\"a}rvel{\"a}}, E., {et~al.} 2018, ArXiv
  e-prints, arXiv:1801.03519

\bibitem[{{Boller} {et~al.}(1996){Boller}, {Brandt}, \&
  {Fink}}]{1996A&A...305...53B}
{Boller}, T., {Brandt}, W.~N., \& {Fink}, H. 1996, \aap, 305, 53

\bibitem[{{Boller} {et~al.}(2016){Boller}, {Freyberg}, {Tr{\"u}mper}, {Haberl},
  {Voges}, \& {Nandra}}]{2016A&A...588A.103B}
{Boller}, T., {Freyberg}, M.~J., {Tr{\"u}mper}, J., {et~al.} 2016, \aap, 588,
  A103

\bibitem[{{Boroson} \& {Green}(1992)}]{1992ApJS...80..109B}
{Boroson}, T.~A., \& {Green}, R.~F. 1992, \apjs, 80, 109

\bibitem[{{Calderone} {et~al.}(2013){Calderone}, {Ghisellini}, {Colpi}, \&
  {Dotti}}]{2013MNRAS.431..210C}
{Calderone}, G., {Ghisellini}, G., {Colpi}, M., \& {Dotti}, M. 2013, \mnras,
  431, 210

\bibitem[{{Chen} {et~al.}(2015){Chen}, {Zhang}, {Zhang}, \&
  {Yu}}]{2015MNRAS.451.4193C}
{Chen}, Y.~Y., {Zhang}, X., {Zhang}, H.~J., \& {Yu}, X.~L. 2015, \mnras, 451,
  4193

\bibitem[{{Congiu} {et~al.}(2017){Congiu}, {Berton}, {Giroletti}, {Antonucci},
  {Caccianiga}, {Kharb}, {Lister}, {Foschini}, {Ciroi}, {Cracco}, {Frezzato},
  {J{\"a}rvel{\"a}}, {La Mura}, {Richards}, \&
  {Rafanelli}}]{2017arXiv170403881C}
{Congiu}, E., {Berton}, M., {Giroletti}, M., {et~al.} 2017, ArXiv e-prints,
  arXiv:1704.03881

\bibitem[{{D'Ammando} {et~al.}(2015){D'Ammando}, {Orienti}, {Larsson}, \&
  {Giroletti}}]{2015MNRAS.452..520D}
{D'Ammando}, F., {Orienti}, M., {Larsson}, J., \& {Giroletti}, M. 2015, \mnras,
  452, 520

\bibitem[{{D'Ammando} {et~al.}(2013){D'Ammando}, {Orienti}, {Doi}, {Giroletti},
  {Dallacasa}, {Hovatta}, {Drake}, {Max-Moerbeck}, {Readhead}, \&
  {Richards}}]{2013MNRAS.433..952D}
{D'Ammando}, F., {Orienti}, M., {Doi}, A., {et~al.} 2013, \mnras, 433, 952

\bibitem[{{Deo} {et~al.}(2006){Deo}, {Crenshaw}, \&
  {Kraemer}}]{2006AJ....132..321D}
{Deo}, R.~P., {Crenshaw}, D.~M., \& {Kraemer}, S.~B. 2006, \aj, 132, 321

\bibitem[{{Doi} {et~al.}(2012){Doi}, {Nagira}, {Kawakatu}, {Kino}, {Nagai}, \&
  {Asada}}]{2012ApJ...760...41D}
{Doi}, A., {Nagira}, H., {Kawakatu}, N., {et~al.} 2012, \apj, 760, 41

\bibitem[{{Doi} {et~al.}(2015){Doi}, {Wajima}, {Hagiwara}, \&
  {Inoue}}]{2015ApJ...798L..30D}
{Doi}, A., {Wajima}, K., {Hagiwara}, Y., \& {Inoue}, M. 2015, \apjl, 798, L30

\bibitem[{{Drake} {et~al.}(2009){Drake}, {Djorgovski}, {Mahabal}, {Beshore},
  {Larson}, {Graham}, {Williams}, {Christensen}, {Catelan}, {Boattini},
  {Gibbs}, {Hill}, \& {Kowalski}}]{2009ApJ...696..870D}
{Drake}, A.~J., {Djorgovski}, S.~G., {Mahabal}, A., {et~al.} 2009, \apj, 696,
  870

\bibitem[{{Fanaroff} \& {Riley}(1974)}]{1974MNRAS.167P..31F}
{Fanaroff}, B.~L., \& {Riley}, J.~M. 1974, \mnras, 167, 31P

\bibitem[{{Gab{\'a}nyi} {et~al.}(2017){Gab{\'a}nyi}, {Frey}, {Paragi},
  {J{\"a}rvel{\"a}}, {Morokuma}, {An}, {Tanaka}, \&
  {Tar}}]{2017arXiv170907202G}
{Gab{\'a}nyi}, K.~{\'E}., {Frey}, S., {Paragi}, Z., {et~al.} 2017, ArXiv
  e-prints, arXiv:1709.07202

\bibitem[{{Giovannini} {et~al.}(2001){Giovannini}, {Cotton}, {Feretti}, {Lara},
  \& {Venturi}}]{2001ApJ...552..508G}
{Giovannini}, G., {Cotton}, W.~D., {Feretti}, L., {Lara}, L., \& {Venturi}, T.
  2001, \apj, 552, 508

\bibitem[{{Gliozzi} {et~al.}(2010){Gliozzi}, {Papadakis}, {Grupe}, {Brinkmann},
  {Raeth}, \& {Kedziora-Chudczer}}]{2010ApJ...717.1243G}
{Gliozzi}, M., {Papadakis}, I.~E., {Grupe}, D., {et~al.} 2010, \apj, 717, 1243

\bibitem[{{Goodrich}(1989)}]{1989ApJ...342..224G}
{Goodrich}, R.~W. 1989, \apj, 342, 224

\bibitem[{{Gu} {et~al.}(2015){Gu}, {Chen}, {Komossa}, {Yuan}, {Shen}, {Wajima},
  {Zhou}, \& {Zensus}}]{2015ApJS..221....3G}
{Gu}, M., {Chen}, Y., {Komossa}, S., {et~al.} 2015, \apjs, 221, 3

\bibitem[{{Hota} {et~al.}(2011){Hota}, {Sirothia}, {Ohyama}, {Konar}, {Kim},
  {Rey}, {Saikia}, {Croston}, \& {Matsushita}}]{2011MNRAS.417L..36H}
{Hota}, A., {Sirothia}, S.~K., {Ohyama}, Y., {et~al.} 2011, \mnras, 417, L36

\bibitem[{{Hota} {et~al.}(2016){Hota}, {Konar}, {Stalin}, {Vaddi}, {Mohanty},
  {Dabhade}, {Dharmik Bhoga}, {Rajoria}, \& {Sethi}}]{2016JApA...37...41H}
{Hota}, A., {Konar}, C., {Stalin}, C.~S., {et~al.} 2016, Journal of
  Astrophysics and Astronomy, 37, 41

\bibitem[{{Ivezi{\'c}} {et~al.}(2002){Ivezi{\'c}}, {Menou}, {Knapp}, {Strauss},
  {Lupton}, {Vanden Berk}, {Richards}, {Tremonti}, {Weinstein}, {Anderson},
  {Bahcall}, {Becker}, {Bernardi}, {Blanton}, {Eisenstein}, {Fan},
  {Finkbeiner}, {Finlator}, {Frieman}, {Gunn}, {Hall}, {Kim}, {Kinkhabwala},
  {Narayanan}, {Rockosi}, {Schlegel}, {Schneider}, {Strateva}, {SubbaRao},
  {Thakar}, {Voges}, {White}, {Yanny}, {Brinkmann}, {Doi}, {Fukugita},
  {Hennessy}, {Munn}, {Nichol}, \& {York}}]{2002AJ....124.2364I}
{Ivezi{\'c}}, {\v Z}., {Menou}, K., {Knapp}, G.~R., {et~al.} 2002, \aj, 124,
  2364

\bibitem[{{Kapahi}(1989)}]{1989AJ.....97....1K}
{Kapahi}, V.~K. 1989, \aj, 97, 1

\bibitem[{{Kellermann} {et~al.}(1989){Kellermann}, {Sramek}, {Schmidt},
  {Shaffer}, \& {Green}}]{1989AJ.....98.1195K}
{Kellermann}, K.~I., {Sramek}, R., {Schmidt}, M., {Shaffer}, D.~B., \& {Green},
  R. 1989, \aj, 98, 1195

\bibitem[{{Kimball} \& {Ivezi{\'c}}(2008)}]{2008AJ....136..684K}
{Kimball}, A.~E., \& {Ivezi{\'c}}, {\v Z}. 2008, \aj, 136, 684

\bibitem[{{Kirk} {et~al.}(2000){Kirk}, {Guthmann}, {Gallant}, \&
  {Achterberg}}]{2000ApJ...542..235K}
{Kirk}, J.~G., {Guthmann}, A.~W., {Gallant}, Y.~A., \& {Achterberg}, A. 2000,
  \apj, 542, 235

\bibitem[{{Komossa} {et~al.}(2006){Komossa}, {Voges}, {Xu}, {Mathur}, {Adorf},
  {Lemson}, {Duschl}, \& {Grupe}}]{2006AJ....132..531K}
{Komossa}, S., {Voges}, W., {Xu}, D., {et~al.} 2006, \aj, 132, 531

\bibitem[{{Konar} \& {Hardcastle}(2013)}]{2013MNRAS.436.1595K}
{Konar}, C., \& {Hardcastle}, M.~J. 2013, \mnras, 436, 1595

\bibitem[{{Konar} {et~al.}(2009){Konar}, {Hardcastle}, {Croston}, \&
  {Saikia}}]{2009MNRAS.400..480K}
{Konar}, C., {Hardcastle}, M.~J., {Croston}, J.~H., \& {Saikia}, D.~J. 2009,
  \mnras, 400, 480

\bibitem[{{Konar} {et~al.}(2013){Konar}, {Hardcastle}, {Jamrozy}, \&
  {Croston}}]{2013MNRAS.430.2137K}
{Konar}, C., {Hardcastle}, M.~J., {Jamrozy}, M., \& {Croston}, J.~H. 2013,
  \mnras, 430, 2137

\bibitem[{{Konar} {et~al.}(2006){Konar}, {Saikia}, {Jamrozy}, \&
  {Machalski}}]{2006MNRAS.372..693K}
{Konar}, C., {Saikia}, D.~J., {Jamrozy}, M., \& {Machalski}, J. 2006, \mnras,
  372, 693

\bibitem[{{Kormendy} \& {Richstone}(1995)}]{1995ARA&A..33..581K}
{Kormendy}, J., \& {Richstone}, D. 1995, \araa, 33, 581

\bibitem[{{L{\"a}hteenm{\"a}ki} {et~al.}(2017){L{\"a}hteenm{\"a}ki},
  {J{\"a}rvel{\"a}}, {Hovatta}, {Tornikoski}, {Harrison}, {L{\'o}pez-Caniego},
  {Max-Moerbeck}, {Mingaliev}, {Pearson}, {Ramakrishnan}, {Readhead}, {Reeves},
  {Richards}, {Sotnikova}, \& {Tammi}}]{2017arXiv170310365L}
{L{\"a}hteenm{\"a}ki}, A., {J{\"a}rvel{\"a}}, E., {Hovatta}, T., {et~al.} 2017,
  ArXiv e-prints, arXiv:1703.10365

\bibitem[{{Laor}(2000)}]{2000ApJ...543L.111L}
{Laor}, A. 2000, \apjl, 543, L111

\bibitem[{{Ledlow} {et~al.}(1998){Ledlow}, {Owen}, \&
  {Keel}}]{1998ApJ...495..227L}
{Ledlow}, M.~J., {Owen}, F.~N., \& {Keel}, W.~C. 1998, \apj, 495, 227

\bibitem[{{Leighly}(1999)}]{1999ApJS..125..317L}
{Leighly}, K.~M. 1999, \apjs, 125, 317

\bibitem[{{Liu} {et~al.}(2016){Liu}, {Yang}, {Supriyanto}, \&
  {Zhang}}]{2016IJAA....6..166L}
{Liu}, X., {Yang}, P., {Supriyanto}, R., \& {Zhang}, Z. 2016, International
  Journal of Astronomy and Astrophysics, 6, 166

\bibitem[{{Liu} {et~al.}(2006){Liu}, {Jiang}, \& {Gu}}]{2006ApJ...637..669L}
{Liu}, Y., {Jiang}, D.~R., \& {Gu}, M.~F. 2006, \apj, 637, 669

\bibitem[{{Mathur} {et~al.}(2001){Mathur}, {Kuraszkiewicz}, \&
  {Czerny}}]{2001NewA....6..321M}
{Mathur}, S., {Kuraszkiewicz}, J., \& {Czerny}, B. 2001, \na, 6, 321

\bibitem[{{Morganti} {et~al.}(1997){Morganti}, {Oosterloo}, {Reynolds},
  {Tadhunter}, \& {Migenes}}]{1997MNRAS.284..541M}
{Morganti}, R., {Oosterloo}, T.~A., {Reynolds}, J.~E., {Tadhunter}, C.~N., \&
  {Migenes}, V. 1997, \mnras, 284, 541

\bibitem[{{Netzer} \& {Trakhtenbrot}(2007)}]{2007ApJ...654..754N}
{Netzer}, H., \& {Trakhtenbrot}, B. 2007, \apj, 654, 754

\bibitem[{{Ohta} {et~al.}(2007){Ohta}, {Aoki}, {Kawaguchi}, \&
  {Kiuchi}}]{2007ApJS..169....1O}
{Ohta}, K., {Aoki}, K., {Kawaguchi}, T., \& {Kiuchi}, G. 2007, \apjs, 169, 1

\bibitem[{{Orr} \& {Browne}(1982)}]{1982MNRAS.200.1067O}
{Orr}, M.~J.~L., \& {Browne}, I.~W.~A. 1982, \mnras, 200, 1067

\bibitem[{{Osterbrock} \& {Pogge}(1985)}]{1985ApJ...297..166O}
{Osterbrock}, D.~E., \& {Pogge}, R.~W. 1985, \apj, 297, 166

\bibitem[{{Paliya} {et~al.}(2018){Paliya}, {Ajello}, {Rakshit}, {Mandal},
  {Stalin}, {Kaur}, \& {Hartmann}}]{2018ApJ...853L...2P}
{Paliya}, V.~S., {Ajello}, M., {Rakshit}, S., {et~al.} 2018, \apjl, 853, L2

\bibitem[{{Rakshit} \& {Stalin}(2017)}]{2017ApJ...842...96R}
{Rakshit}, S., \& {Stalin}, C.~S. 2017, \apj, 842, 96

\bibitem[{{Rakshit} {et~al.}(2017){Rakshit}, {Stalin}, {Chand}, \&
  {Zhang}}]{2017ApJS..229...39R}
{Rakshit}, S., {Stalin}, C.~S., {Chand}, H., \& {Zhang}, X.-G. 2017, \apjs,
  229, 39

\bibitem[{{Rani} {et~al.}(2017){Rani}, {Stalin}, \&
  {Rakshit}}]{2017MNRAS.466.3309R}
{Rani}, P., {Stalin}, C.~S., \& {Rakshit}, S. 2017, \mnras, 466, 3309

\bibitem[{{Richards} \& {Lister}(2015)}]{2015ApJ...800L...8R}
{Richards}, J.~L., \& {Lister}, M.~L. 2015, \apjl, 800, L8

\bibitem[{{Sesar} {et~al.}(2007){Sesar}, {Ivezi{\'c}}, {Lupton}, {Juri{\'c}},
  {Gunn}, {Knapp}, {DeLee}, {Smith}, {Miknaitis}, {Lin}, {Tucker}, {Doi},
  {Tanaka}, {Fukugita}, {Holtzman}, {Kent}, {Yanny}, {Schlegel}, {Finkbeiner},
  {Padmanabhan}, {Rockosi}, {Bond}, {Lee}, {Stoughton}, {Jester}, {Harris},
  {Harding}, {Brinkmann}, {Schneider}, {York}, {Richmond}, \& {Vanden
  Berk}}]{2007AJ....134.2236S}
{Sesar}, B., {Ivezi{\'c}}, {\v Z}., {Lupton}, R.~H., {et~al.} 2007, \aj, 134,
  2236

\bibitem[{{Singal}(1993)}]{1993MNRAS.263..139S}
{Singal}, A.~K. 1993, \mnras, 263, 139

\bibitem[{{Singh} {et~al.}(2015){Singh}, {Ishwara-Chandra}, {Wadadekar},
  {Beelen}, \& {Kharb}}]{2015MNRAS.446..599S}
{Singh}, V., {Ishwara-Chandra}, C.~H., {Wadadekar}, Y., {Beelen}, A., \&
  {Kharb}, P. 2015, \mnras, 446, 599

\bibitem[{{Sulentic} {et~al.}(2000){Sulentic}, {Zwitter}, {Marziani}, \&
  {Dultzin-Hacyan}}]{2000ApJ...536L...5S}
{Sulentic}, J.~W., {Zwitter}, T., {Marziani}, P., \& {Dultzin-Hacyan}, D. 2000,
  \apjl, 536, L5

\bibitem[{{Tanaka} {et~al.}(2014){Tanaka}, {Morokuma}, {Itoh}, {Akitaya},
  {Tominaga}, {Saito}, {Stawarz}, {Tanaka}, {Gandhi}, {Ali}, {Aoki},
  {Contreras}, {Doi}, {Essam}, {Hamed}, {Hsiao}, {Iwata}, {Kawabata}, {Kawai},
  {Kikuchi}, {Kobayashi}, {Kuroda}, {Maehara}, {Matsumoto}, {Mazzali},
  {Minezaki}, {Mito}, {Miyata}, {Miyazaki}, {Mori}, {Moritani},
  {Morokuma-Matsui}, {Morrell}, {Nagao}, {Nakada}, {Nakata}, {Noma}, {Ohsuga},
  {Okada}, {Phillips}, {Pian}, {Richmond}, {Sahu}, {Sako}, {Sarugaku},
  {Shibata}, {Soyano}, {Stritzinger}, {Tachibana}, {Taddia}, {Takaki}, {Takey},
  {Tarusawa}, {Ui}, {Ukita}, {Urata}, {Walker}, \&
  {Yoshii}}]{2014ApJ...793L..26T}
{Tanaka}, M., {Morokuma}, T., {Itoh}, R., {et~al.} 2014, \apjl, 793, L26

\bibitem[{{Tempel} {et~al.}(2011){Tempel}, {Saar}, {Liivam{\"a}gi}, {Tamm},
  {Einasto}, {Einasto}, \& {M{\"u}ller}}]{2011A&A...529A..53T}
{Tempel}, E., {Saar}, E., {Liivam{\"a}gi}, L.~J., {et~al.} 2011, \aap, 529, A53

\bibitem[{{V{\'e}ron-Cetty} {et~al.}(2001){V{\'e}ron-Cetty}, {V{\'e}ron}, \&
  {Gon{\c c}alves}}]{2001A&A...372..730V}
{V{\'e}ron-Cetty}, M.-P., {V{\'e}ron}, P., \& {Gon{\c c}alves}, A.~C. 2001,
  \aap, 372, 730

\bibitem[{{Wang} {et~al.}(1996){Wang}, {Brinkmann}, \&
  {Bergeron}}]{1996A&A...309...81W}
{Wang}, T., {Brinkmann}, W., \& {Bergeron}, J. 1996, \aap, 309, 81

\bibitem[{{Whalen} {et~al.}(2006){Whalen}, {Laurent-Muehleisen}, {Moran}, \&
  {Becker}}]{2006AJ....131.1948W}
{Whalen}, D.~J., {Laurent-Muehleisen}, S.~A., {Moran}, E.~C., \& {Becker},
  R.~H. 2006, \aj, 131, 1948

\bibitem[{{Wright} {et~al.}(2010){Wright}, {Eisenhardt}, {Mainzer}, {Ressler},
  {Cutri}, {Jarrett}, {Kirkpatrick}, {Padgett}, {McMillan}, {Skrutskie},
  {Stanford}, {Cohen}, {Walker}, {Mather}, {Leisawitz}, {Gautier}, {McLean},
  {Benford}, {Lonsdale}, {Blain}, {Mendez}, {Irace}, {Duval}, {Liu}, {Royer},
  {Heinrichsen}, {Howard}, {Shannon}, {Kendall}, {Walsh}, {Larsen}, {Cardon},
  {Schick}, {Schwalm}, {Abid}, {Fabinsky}, {Naes}, \&
  {Tsai}}]{2010AJ....140.1868W}
{Wright}, E.~L., {Eisenhardt}, P.~R.~M., {Mainzer}, A.~K., {et~al.} 2010, \aj,
  140, 1868

\bibitem[{{Yao} {et~al.}(2015){Yao}, {Yuan}, {Zhou}, {Komossa}, {Zhang},
  {Qiao}, \& {Liu}}]{2015MNRAS.454L..16Y}
{Yao}, S., {Yuan}, W., {Zhou}, H., {et~al.} 2015, \mnras, 454, L16

\bibitem[{{Yuan} {et~al.}(2008){Yuan}, {Zhou}, {Komossa}, {Dong}, {Wang}, {Lu},
  \& {Bai}}]{2008ApJ...685..801Y}
{Yuan}, W., {Zhou}, H.~Y., {Komossa}, S., {et~al.} 2008, \apj, 685, 801

\bibitem[{{Zhou} {et~al.}(2006){Zhou}, {Wang}, {Yuan}, {Lu}, {Dong}, {Wang}, \&
  {Lu}}]{2006ApJS..166..128Z}
{Zhou}, H., {Wang}, T., {Yuan}, W., {et~al.} 2006, \apjs, 166, 128

\end{thebibliography}

\end{document}